\documentclass[useAMS,usenatbib]{mn2e}

\usepackage{fixltx2e,ifpdf,amsmath,amssymb,nicefrac}
\usepackage[multiple]{footmisc}
\ifpdf
  \usepackage[pdftex]{graphicx,color}
\else
  \usepackage[dvips]{graphicx,color}
\fi

\newcommand{\Teff}{\ensuremath{T_{\rm eff}}}
\newcommand{\logg}{\ensuremath{\log{g}}}

\newcommand{\vt}{\ensuremath{v_{\rm t}}}
\newcommand{\lH}{\ensuremath{\ell/H}}
\newcommand{\MH}{\ensuremath{\mbox{[M/H]}}}
\newcommand{\kms} {\mbox{\rm km$\;$s$^{-1}$}}

\title[Limb darkening and exotransits]
{On stellar limb darkening and exoplanetary transits.}
\author
[Ian D. Howarth]
{Ian D. Howarth\thanks{E-mail: idh@star.ucl.ac.uk}\\
Dept.\ Physics  \& Astronomy, UCL, Gower Street, London
WC1E~6BT, UK}
\begin{document}

\date{}

\pagerange{\pageref{firstpage}--\pageref{lastpage}} \pubyear{2011}

\maketitle

\label{firstpage}

\begin{abstract}
  This paper examines how to compare stellar limb-darkening
  coefficients evaluated from model atmospheres with those derived
  from photometry.  Different characterizations of a given model
  atmosphere can give quite different numerical results (even for a
  given limb-darkening `law'), while light-curve analyses yield
  limb-darkening coefficients that are dependent on system geometry,
  and that are not directly comparable to any model-atmosphere
  representation.  These issues are examined in the context of
  exoplanetary transits, which offer significant advantages over
  traditional binary-star eclipsing systems in the study of stellar
  limb darkening.  `Like for like' comparisons between light-curve
  analyses and new model-atmosphere results, mediated by synthetic
  photometry, are conducted for a small sample of stars.  Agreement
  between the resulting synthetic-photometry/atmosphere-model (SPAM)
  limb-darkening coefficients and empirical values ranges from very
  good to quite poor, even though the targets investigated show only a
  small dispersion in fundamental stellar parameters.
\end{abstract}


\section{Introduction}

Stellar limb darkening is the wavelength-dependent decrease in specific intensity,
$I_{\lambda}(\mu)$, with decreasing $\mu$, where $\mu =\cos\theta$ and
$\theta$ is the angle between the surface normal and the line of
sight;\footnote{Under some circumstances, limb \textit{brightening}
  can occur.}\footnote{The specific intensity,
referred to as the
radiance in other contexts,
is
the rate of energy flow
per unit area,
per unit time,
per unit wavelength interval,
per unit solid angle.  Expressing $I_\lambda$ as a function of a single
angle $\mu$ makes the implicit assumption of azimuthal symmetry of the
radiation field.}   
in the context of model atmospheres, it is,
in principle,  significantly more
sensitive to input physics than are integral quantities,
such as the emergent flux.

Until rather recently, the only important
opportunity to compare models and observations 
of limb darkening for the distant stars
has been through eclipsing-binary systems, but there the comparison
has been hindered both by the
rather weak dependence on limb darkening of the light-curves, and by degeneracies with
other model parameters.  As a
consequence, normal practice among light-curve
analysts has been to \textit{assume} some description of limb
darkening, based on stellar-atmosphere results; any errors in the
description are liable to be concealed by small adjustments to 
fitted free
parameters.

New observational techniques have begun to allow the direct
investigation of limb darkening (and hence more sensitive tests of
model-atmosphere calculations) under other circumstances.  Optical
interferometry has opened the way to direct imaging of stellar
surfaces beyond the solar system for a handful of stars with the
largest angular diameters \citep[e.g.,][]{Aufdenberg05}, and microlensing
light-curves are also capable of probing the intensity distribution of
the lensed source \citep[e.g.,][]{Witt95, Zub11},
albeit usually only crudely \citep{Dominik04}.  However, the
focus of the present paper is on the role of limb darkening in
exoplanetary transits, which are likely to yield many more results in
the coming years than any other technique.

In many respects, star+exoplanet systems are close to being idealised
eclipsing binaries: it is often reasonable to assume that the
photo\-metric properties of the parent star are unaffected by the
transiting planet (i.e., no tidal distortion, `reflection' effect, or
gravity darkening), and that the secondary (planet) is completely
dark, and spherical.  These assumptions reduce the number of geometric
unknowns to be determined from the light-curve to only three (in
addition to the orbital ephemeris, which may be established
separately); e.g., the ratio of the radii, the size of the star in
units of the centres-of-mass separation, and the impact
parameter.  This relative simplicity allows a more critical
examination of limb darkening than is possible in star+star systems.
With an anticipated torrent of data of extremely high quality from
satellites such as \textit{Kepler}, it is therefore timely to revisit
the comparison of limb-darkening coefficients (LDCs) from
model-atmosphere and light-curve analyses, as has already been
recognized by several authors \citep[e.g.][]{Southworth08, Pal08,
  Claret09}.

This comparison is examined here as follows: Section~\ref{sec:CLD}
reviews limb-darkening `laws' and fitting techniques (including a new
flux-conserving least-squares methodology), stressing the spread in
numerical coefficients that can arise even when characterizing a given
model-atmosphere intensity distribution with a given law.
Section~\ref{sec:EXOX} examines the LDCs extracted from light-curve
analyses, emphasizing not only the range in numerical coefficients that can
arise from characterizing a given surface-brightness distribution
under different geometries, but also that the photometrically determined
LDCs are not, in any case, directly comparable to those derived from
model-atmosphere calculations.  

With the background that (i) the numerical values of coefficients
determined from model atmospheres depend on the fitting method, and
(ii) coefficients determined from light-curves are not directly
comparable to model-atmosphere resulst, and vary with impact
parameter, Section~\ref{sec:LfL} outlines how model-atmosphere
limb-darkening results \textit{can} be compared with inferences from
high-quality exotransit photometry, and presents illustrative results.

\section{Characterizing limb darkening}
\label{sec:CLD}

In photo\-metric analyses, it is still impractical to invert observed
light-curves in order to recover detailed stellar surface-brightness
distributions.  Rather, in this context limb darkening is habitually
represented by some \textit{ad hoc} `law' with one or, at most, two
free parameters, which may be optimized as part of the fitting
process.\footnote{In practice, even `two-parameter' fits still allow
  only one coefficient to be usefully constrained; see the discussion
  in Section~\ref{sec:LfL}.}  In order to facilitate comparison of
light-curve results with model-atmosphere calculations,
model
intensities are often represented with the same parametrizations.

\subsection{Functional forms}

Historically, the first limb-darkening law to be developed
was the analytical solution for an atmosphere in which the source
function is linear in optical depth:
\begin{align}
I_{\lambda}(\mu)  = I_\lambda(1)\left[{1 - u(1-\mu)}\right]
\label{eq:01}
\end{align}
\citep{Schwarz06}, where the wavelength dependence of $u$ is implicit
(although $u=0.6$ at all wavelengths for a grey atmosphere; 
\citealt{Milne21}).
This linear law is the universally
adopted one-parameter representation of limb darkening.

More-realistic atmosphere models do not have analytical functional
representations of actual limb darkening.  Following the work of
\citet{Kopal49}, a quadratic law of the form
\begin{align}
I(\mu) = I(1)\left[{1 - u_1(1-\mu) - u_2(1-\mu)^2}\right]
\label{eq:02}
\end{align}
has been widely adopted as a characterization of model-atmosphere
calculations.  It is of particular importance in modelling
exotransit photometry using Monte-Carlo Markov-Chain (MCMC)
techniques, since it allows for analytical calculation of light-curves
with good computational efficiency \citep{Mandel02}.

While eqtns.~\ref{eq:01} and \ref{eq:02} are convenient in the analysis 
of light-curves,\footnote{A number of other limb-darkening laws have been proposed;
cf., e.g., \citet{Diaz92}}
a
significantly more accurate representation of model-atmosphere results
is achieved with the four-coefficient fit introduced by
\citet{Claret00}:
\begin{align}
I(\mu) = I(1)\left[{1 -
\sum_{n=1}^4a_n\left({1 - \mu^{n/2}}\right)}\right].
\label{eq:03}
\end{align}
This form reproduces intensities from model atmospheres to 
$\sim$1 part in 1000 over a wide range of parameter space
\citep[e.g.,][]{Howarth11}, although it isn't practical to estimate
numerical values of the coefficients from photometry.

\subsection{Fitting model-atmosphere intensities.}
\label{CMA}

Although linear and quadratic limb-darkening laws may not give
particularly accurate functional descriptions of model-atmosphere
intensities, it is nonetheless necessary to represent them in this way in
order to compare with observationally derived LDCs.
However, even for a given limb-darkening
law, the characterization of model-atmosphere results 
using different
fitting techniques
can result in
quite different values for the coefficients 

\subsubsection{LS1: least squares with $I(1)$ constrained}
Rewriting eqtn.~\ref{eq:01} as
\begin{align*}
I_{\lambda}(\mu)/I_\lambda(1)  = \left[{1 - u(1-\mu)}\right],
\end{align*}
gives a one-parameter formulation straightforwardly solved by least
squares for $u$, using as input the model-atmosphere values of $I(\mu)$.  The
intercept of the linear fit is implicitly constrained such that
$\hat{I}(1)$, the value of $I(1)$ evaluated from the fitted law, is
fixed at model-atmosphere value.   The quadratic equivalent is
\begin{align*}
I_{\lambda}(\mu)/I_\lambda(1)  = \left[{1 - u_1(1-\mu) - u_2(1-\mu)^2}\right],
\end{align*}

\subsubsection{LS2: least squares with $I(1)$ free}
Relaxing the constraint that
$\hat{I}_\lambda(1) \equiv I_\lambda(1)$ gives laws that can be
again be solved in a trivial least-squares exercise,
with
$\hat{I}_\lambda(1)$ as an additional free parameter:
\begin{align*}
I_{\lambda}(\mu)  = \hat{I}_\lambda(1)\left[{1 - u(1-\mu)}\right]
\end{align*}
(linear),\goodbreak
\begin{align*}
I_{\lambda}(\mu)  = \hat{I}_\lambda(1)\left[{1 - u_1(1-\mu) - u_2(1-\mu)^2}\right]
\end{align*}
(quadratic;  in practice, both sides may be divided by $I_\lambda(1)$
in these two equations).

\subsubsection{Flux-conserving fit:  FC1}

The physical flux $F_\lambda$ 
is related to the specific intensity through
\begin{align}
\begin{split}
F_\lambda &= 2\pi \int_0^1{I_\lambda(\mu) \mu \,\mbox{d}\mu}\\
& = 4 \pi H_\lambda
\end{split}
\label{eq:flux}
\end{align}
where $H_\lambda$ is the Eddington flux (the first-order moment of the
radiation field).
The integration of  eqtn.~\ref{eq:flux}
using an analytical limb-darkening law to represent $I_\lambda(\mu)$,
with coefficients 
determined by least squares, will not normally recover the physical
flux exactly.   To address this, we can impose the condition that
\begin{align*}
F_\lambda &= 2\pi \int_0^1{\hat{I}_\lambda(\mu) \mu \,\mbox{d}\mu};\\
\intertext{that is,}
F_\lambda &= \pi\hat{I}(1)\left[{1 - u/3}\right]\\
F_\lambda &= \frac{2\pi \hat{I}(1)}{12}\left[{6 - 2u_1 - u_2}\right]
\end{align*}
in the linear and quadratic cases, respectively.
Requiring $\hat{I}_\lambda(\mu)$, evaluated from the limb-darkening law,
to equal $I_\lambda(\mu)$, evaluated from the model atmosphere, at some arbitrary $\mu
= x$, we obtain
\begin{align}
u = \frac{\pi I_\lambda(x) - F_\lambda}{(\pi I_\lambda(x)/3) +
  (x - 1)F_\lambda}.
\end{align}
for the linear law.   \citet{Wade85} chose $x=1$, whence
\begin{align*}
u = 3\left[{ 1 - F_\lambda/\left({ \pi I_\lambda(1)}\right)    }\right]
\end{align*}
(noting the \citeauthor{Wade85}'s ``angle-averaged'' [astrophysical]
flux is $F_\lambda/\pi$ in the nomenclature adopted here).  In effect, the
choice of $x$ fixes the intercept of the linear law, with the
constraint of flux conservation then fixing the slope.  

The equivalent
algebra for the quadratic law follows from selecting any two values
$\mu = x_1, x_2$ at which $\hat{I}_\lambda(\mu)$ is equal to $I_\lambda(\mu)$, giving
a pair of simultaneous equations that can readily be solved for $u_1,
u_2$.  \citet{Wade85}, and subsequent authors, used $x_1 = 1, x_2 = 0.1$ (values which are also
adopted here), but again these are more or less arbitrary choices.

\begin{figure}
\center{\includegraphics[scale=0.3,angle=270]{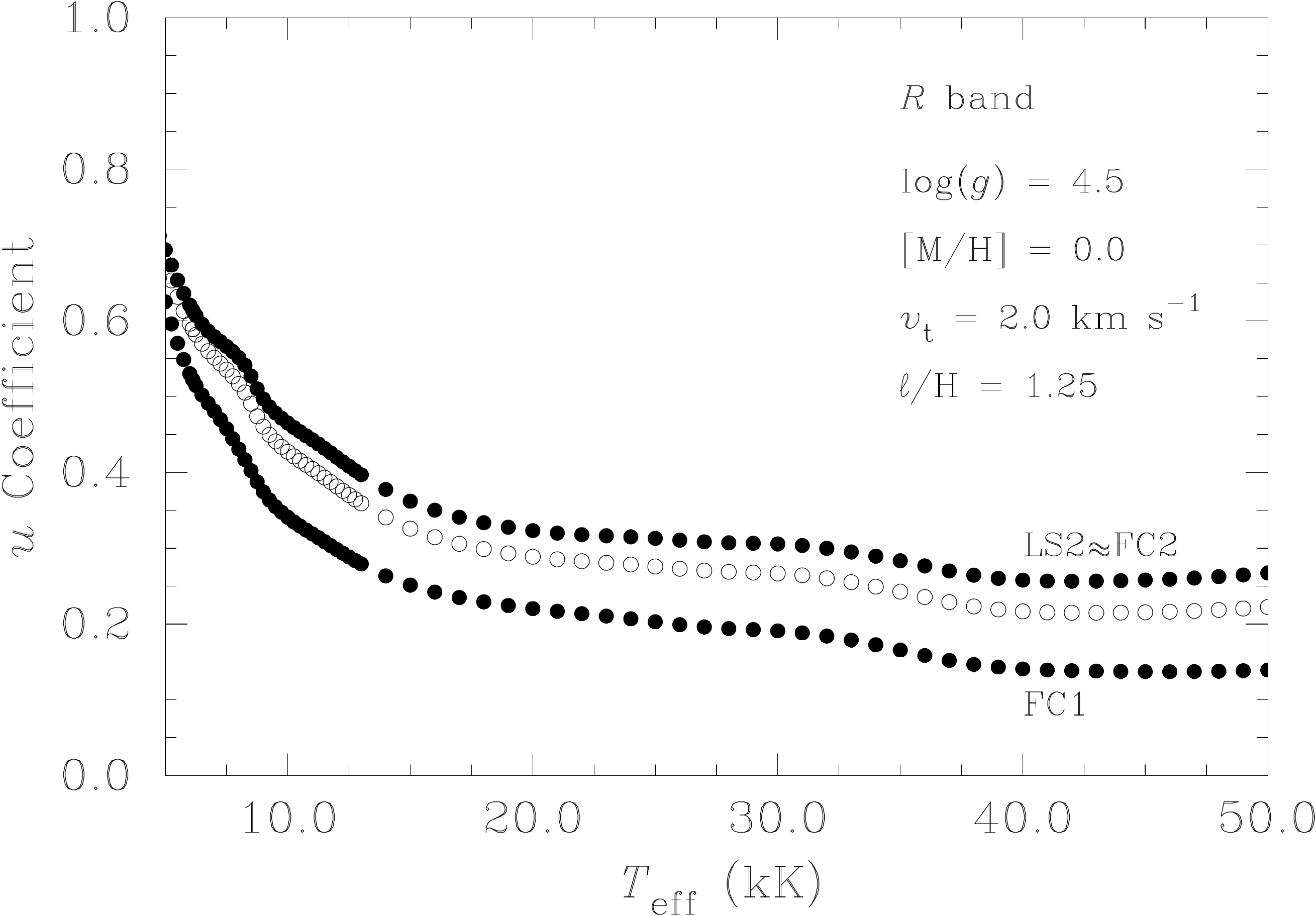}} 
\caption{Comparison of linear limb-darkening coefficients determined
from model atmospheres  by different numerical techniques (cf.\
Section~\ref{sec:CLD}), as a function of effective temperature.  Open
circles are LS1 results.}
\label{fig:Rcomp}
\end{figure}

\begin{figure}
\center{\includegraphics[scale=0.73,angle=0]{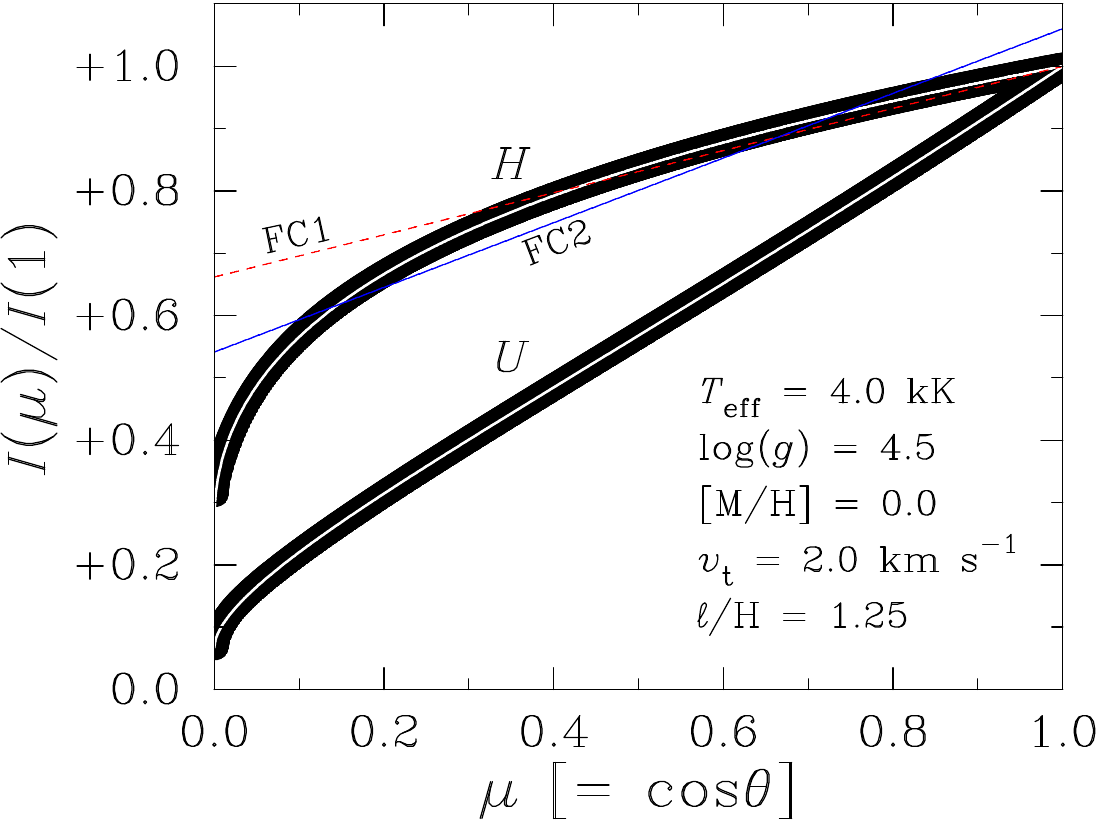}} 
\caption{Limb darkening 
  in the \textit{H} and \textit{U} bands for a 4~kK model.
  The thick `lines' are individual model-atmosphere intensities, shown
  as points which
  merge together at this scale.   The fitted 4-coefficient
  limb-darkening laws are shown drawn through the points.  Straight
  lines show the linear limb-darkening laws for the \textit{H} band,  with coefficients
  determined by standard flux conservation and by flux-conserving least
  squares
(FC1, FC2, respectively; cf.~Section~\ref{CMA}).}
\label{fig:uPlot}
\end{figure}

\subsubsection{Flux-conserving least squares:  FC2}

The weakness of the standard flux-conserving approach is the lack of a
compelling physical argument to select any particular $x$ values for
the normalization (other than requiring the intensities to be
everywhere positive; e.g., requiring
$0 \geq u \geq 1$ in the linear case). 

Rather than making an arbitrary choice of $x$, we can instead
introduce the more objective requirement of minimising the sum of the
squares of the differences between model and fitted intensities while
still requiring flux to be conserved.  For a linear law it is
convenient first to determine $u$ by 
by minimising
$\sum{\left({\hat{I}(\mu) - I(\mu)}\right)^2}$, 
using standard least-squares techniques, 
where
\begin{align*}
\hat{I}_\lambda(\mu) = \frac{3F_\lambda}{\pi}\left[{
\frac{1 - u(1-\mu)}{3 - u}}\right];
\end{align*}
and to then evaluate
\begin{align*}
\hat{I}_\lambda(1) = \frac{3F_\lambda}{\pi(3-u)}.
\end{align*}
Corresponding results for the quadratic law are
\begin{align*}
\hat{I}_\lambda(\mu) = \frac{6F_\lambda}{\pi}\left[{
\frac{1 - u_1(1-\mu) - u_2(1-\mu)^2}{6 - 2u_1 - u_2}}\right]
\end{align*}
\begin{align*}
\hat{I}_\lambda(1) = \frac{6F_\lambda}{\pi(6 - 2u_1 - u_2)}.
\end{align*}
Not surprisingly, this newly introduced approach of flux-conserving
least squares generally yields numerical coefficients very close
to those found using the LS2 method.   Therefore, although it may be regarded as
superior to LS2 in principle, in practice it affords no great benefit
(and turns out not to give results particularly close to photometrically
inferred LDCs).

\subsection{Other numerical factors}
\label{sec:onf}

The foregoing numerical methods can (and do) yield substantially
different LDC values, even for the standard linear and quadratic
representations of a given, fixed, intensity distribution, as is
illustrated by Figs.~\ref{fig:Rcomp} and~\ref{fig:uPlot}.  
For
a given intensity distribution, in the optical wavelength regime the
FC1 $u$ coefficient is usually the smallest numerically; LS2 and FC2
$u$ coefficients are very similar, and relatively large; and the
LS1 coefficient is intermediate.

When characterizing model-atmosphere results, the density and
distribution of angles at which intensities are calculated, and the
weighting scheme, will also influence the numerical values of
limb-darkening coefficients \citep[e.g.,][]{Diaz92, Claret08}.  In the
present work, intensities were computed for $\mu = 0.001$ to 1 at
steps of 0.001,\footnote{The intensities were calculated in detail,
  not `densified' from a sparser grid.} and equally weighted when
fitting functional forms.

Of course, the physics used in constructing the model atmosphere is
also critical.  The limb-darkening coefficients used throughout this
paper were computed using the {\sc Atlas9} line-blanketed LTE
model-atmosphere code \citep{Kurucz93}, as ported to {\sc gnu}-linux
systems by \citet{Sbordone07}, with the Opacity Distribution Functions
described by \citet{Howarth11}. Solar abundances, a microturbulent
velocity of $\vt = 2$~\kms, and mixing-length parameter $\ell/H =
1.25$ were adopted unless noted otherwise.  These models use
time-independent, plane-parallel structures; while atmospheric
extension is unlikely to be important in the parameter space discussed
here, the neglect of time-dependent 3D effects may be significant when
comparing with empirical results (e.g., \citealt{Bigot06}).

\begin{figure*}
\center{\includegraphics[scale=0.8,angle=0]{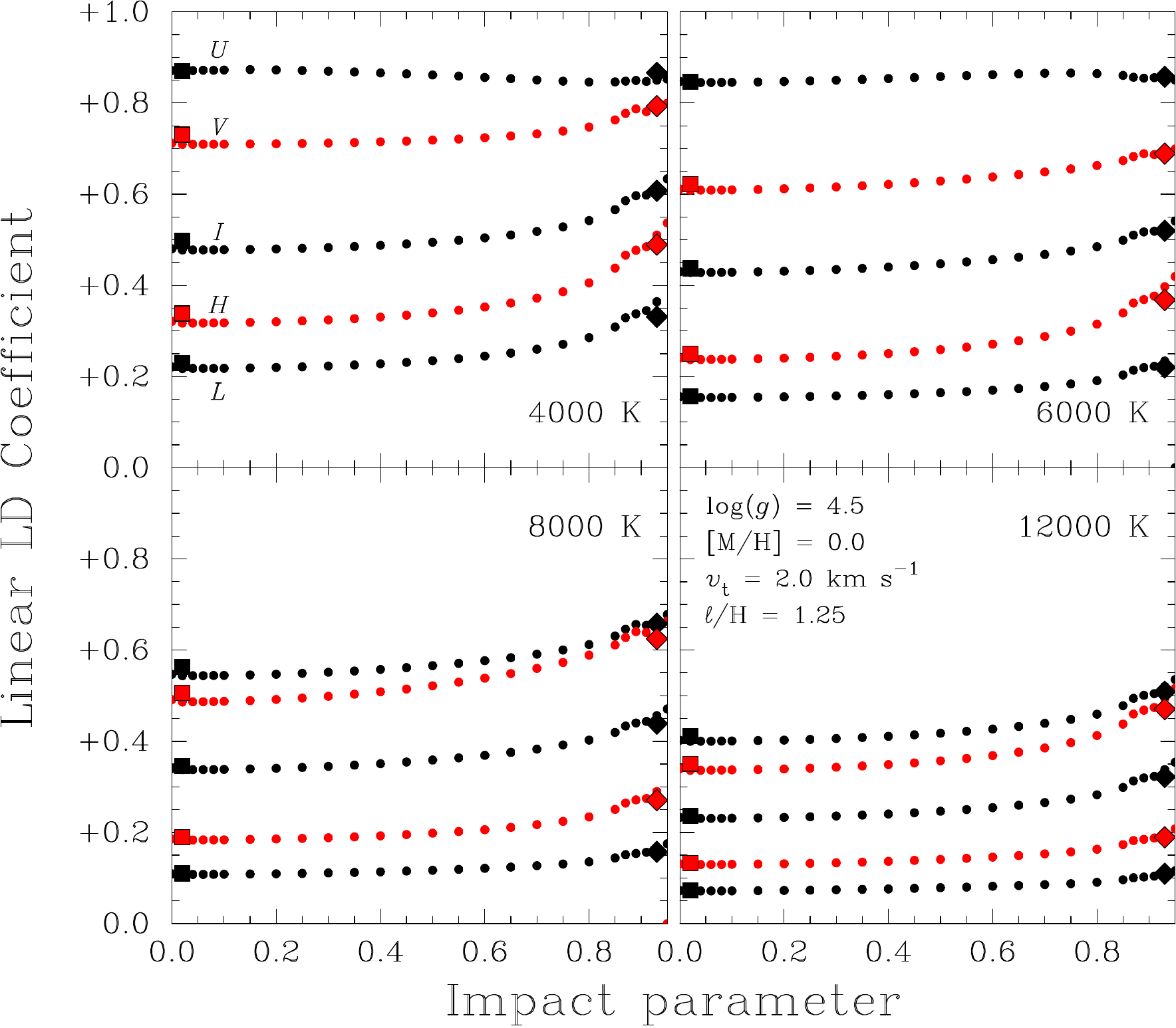}} 
\caption{Photo\-metrically determined linear limb-darkening coefficients
  (small dots), for fixed input limb darkening,
  in the Johnson-Cousins-Glass \textit{UVIHL} passbands
(cf.~Section~\ref{sec:llx}).  Larger symbols show
  corresponding linear limb-darkening coefficients determined directly
  from the \textit{same} input model-atmosphere intensity distributions using
flux-conserving and flux-conserving least-squares fitting
(FC1, large squares, FC2, large diamonds; Section~\ref{CMA}. These single-valued results are plotted at arbitrary 
  impact parameters).}
\label{fig:uComp}
\end{figure*}

\begin{figure*}
\center{\includegraphics[scale=0.73,angle=0]{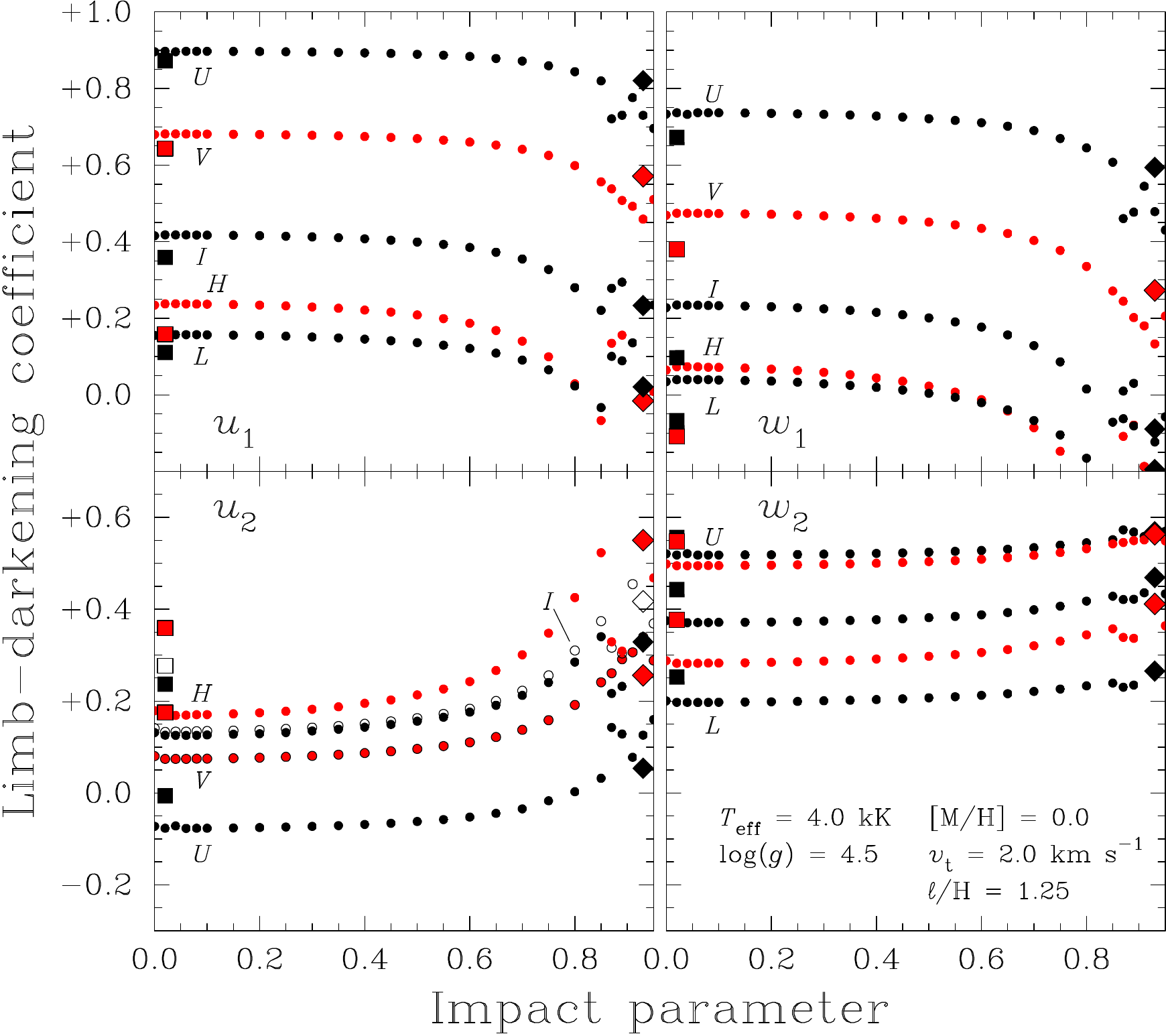}} 
\caption{Photo\-metrically determined quadratic limb-darkening
  coefficients for a 4~kK model in the Johnson-Cousins-Glass
  \textit{UVIHL} passbands.  Left-hand panels, $u_1, u_2$ coefficients
  (eqtn.~\ref{eq:02}); right-hand panels, rotated $w_1, w_2$
  coefficients (Section~\ref{sec:rot1}).  
Larger dots show corresponding quadratic
  limb-darkening coefficients determined directly from the
  \textit{same} input model-atmosphere intensity distributions using
  flux-conserving and flux-conserving least-squares fitting (FC1,
  large squares, FC2, large diamonds; Section~\ref{CMA}. These
  single-valued results are plotted at arbitrary impact parameters).}
\label{fig:qPlot}
\end{figure*}

\begin{figure*}
\center{\includegraphics[scale=0.73,angle=0]{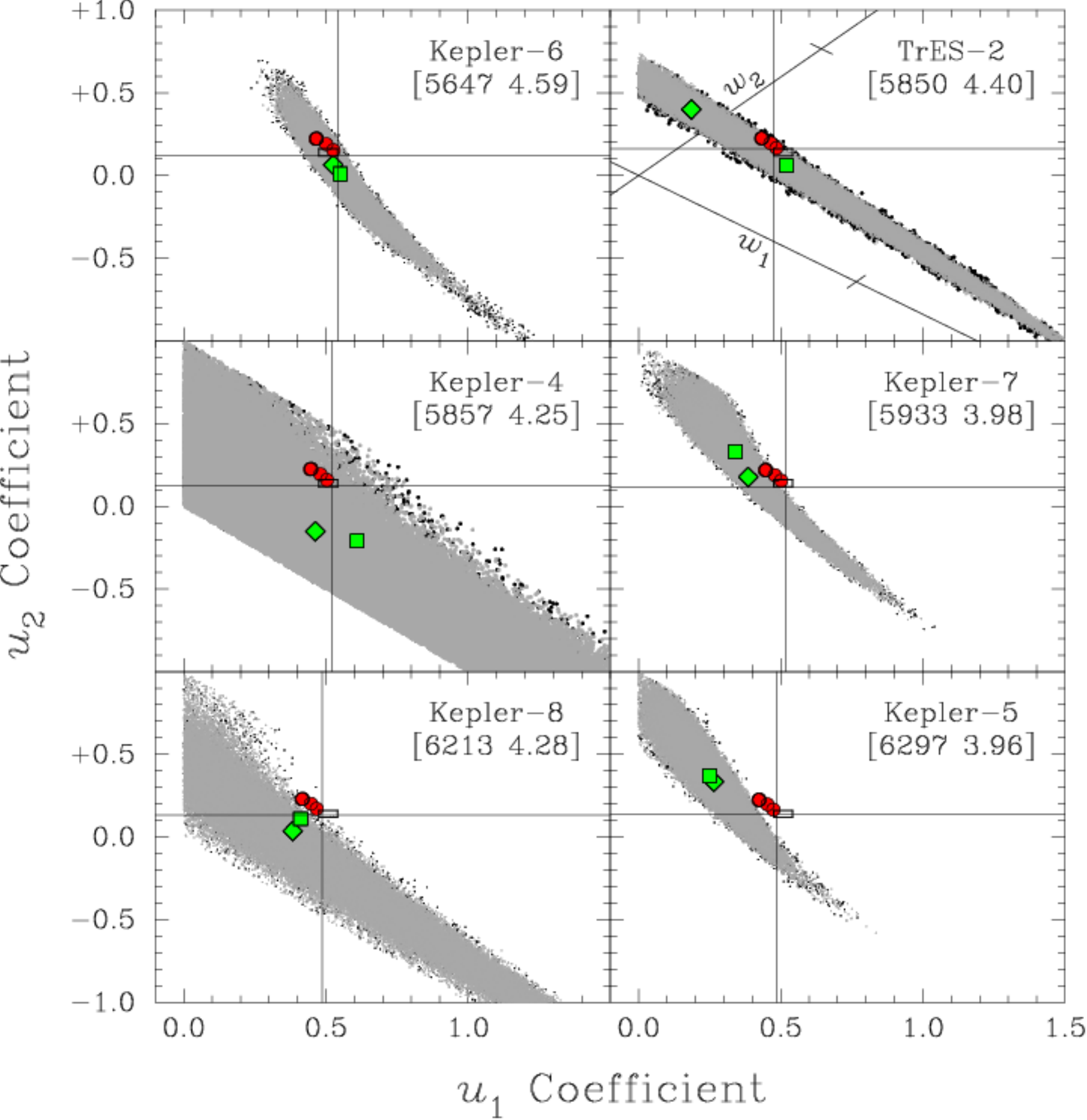}} 
\caption[]{Quadratic limb-darkening coefficients determined by
  \citet{Kipping11a, Kipping11b} from \textit{Kepler} photometry
  compared with model-atmosphere results.  Panels are identified by
  star name, \Teff, and \logg.  Bands of grey points show the
  projection onto the ($u_1, u_2$) plane of the 90\%\ of MCMC results
  yielding the smallest $\chi^2$ values, which overlay the best 95\%\
  results (black points, not visible in all frames because this figure
  is a projection of multiparameter modelling onto a specific 2D
  plane).  
  Green squares show the median values from MCMC runs (the solutions
adopted by \citealt{Kipping11a, Kipping11b}), and
green diamonds the minimum-$\chi^2$ MCMC results.
  Red
  dots show fits to model-atmosphere intensities (left to right:
  FC2/LS2 [indistinguishable at this scale], LS1, FC1), while
  horizontal and vertical lines indicate the SPAM solutions.  The
  small rectangle shown in each panel (perhaps most easily seen by
  zooming in on the on-line version) encompasses the SPAM solutions
  for all six targets, and is included to provide an qualitative
  indication of the rather small scale of uncertainties likely to
  result from any plausible errors in input stellar parameters.
  The rotated $w_1, w_2$ axes (eqtn.~\ref{eq:w1w2}) are shown in the
  TrES-2 panel, for reference; by design, most of the variance in the
  MCMC results is in $w_1$.  }
\label{fig:P6u}
\end{figure*}


\begin{figure}
\center{\includegraphics[scale=0.6,angle=270]{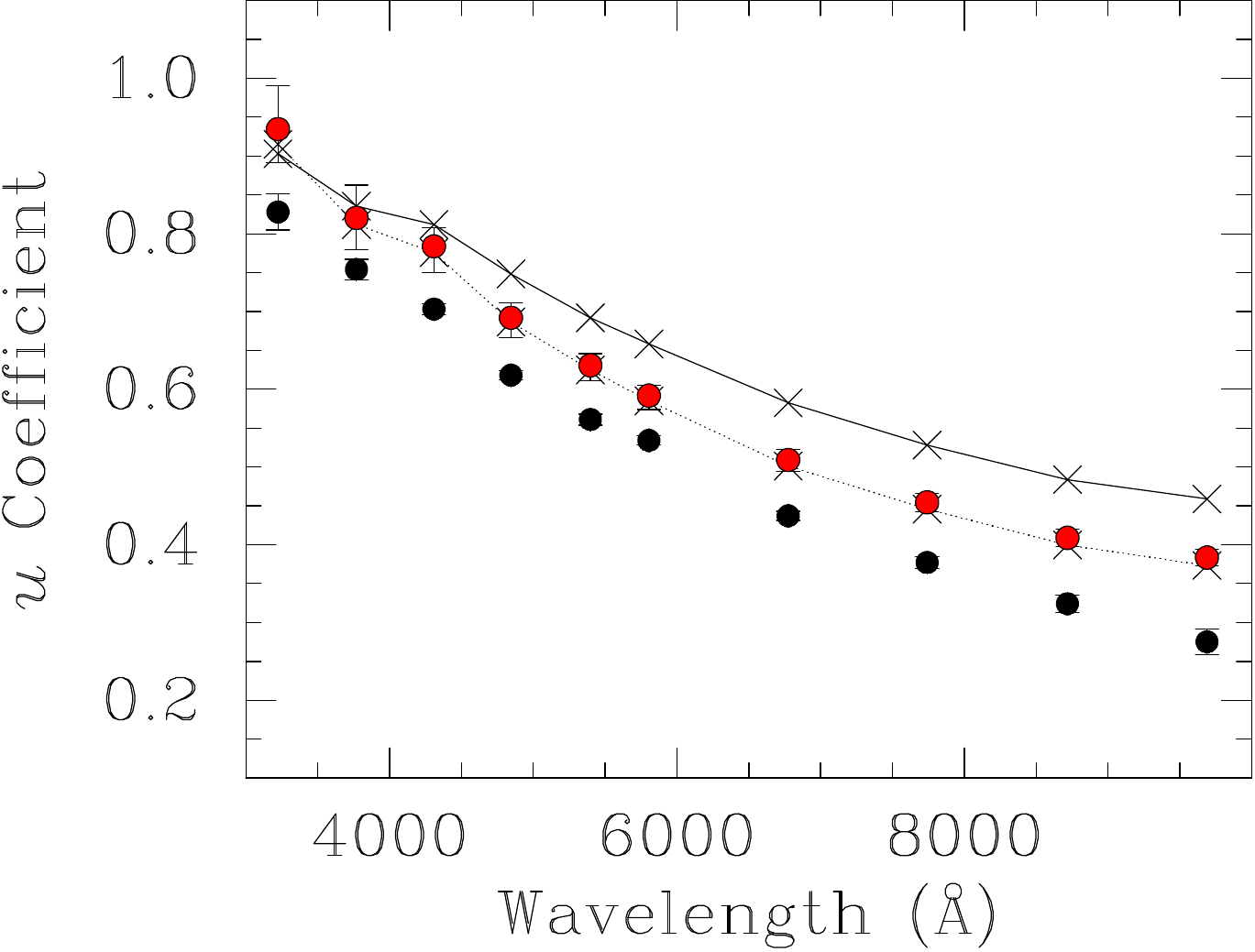}} 
\caption{Comparison of linear limb-darkening coefficients determined
  observationally for HD~209458 (lower set of black points, from
  \citealt{Southworth08}) and SPAM calculations (upper set of red
  points).  The `error bars' on the model-atmosphere results are the
  result of varying input stellar parameters (see
  Section~\ref{sec:HDX} for details); for both SPAM and empirical
  results, the error bars are smaller than the points at most
  wavelengths.  Continuous and dotted lines connect LS2 and FC1
  model-atmosphere results, respectively.}
\label{fig:HD1}
\end{figure}

\begin{figure}
\center{\includegraphics[scale=0.6,angle=270]{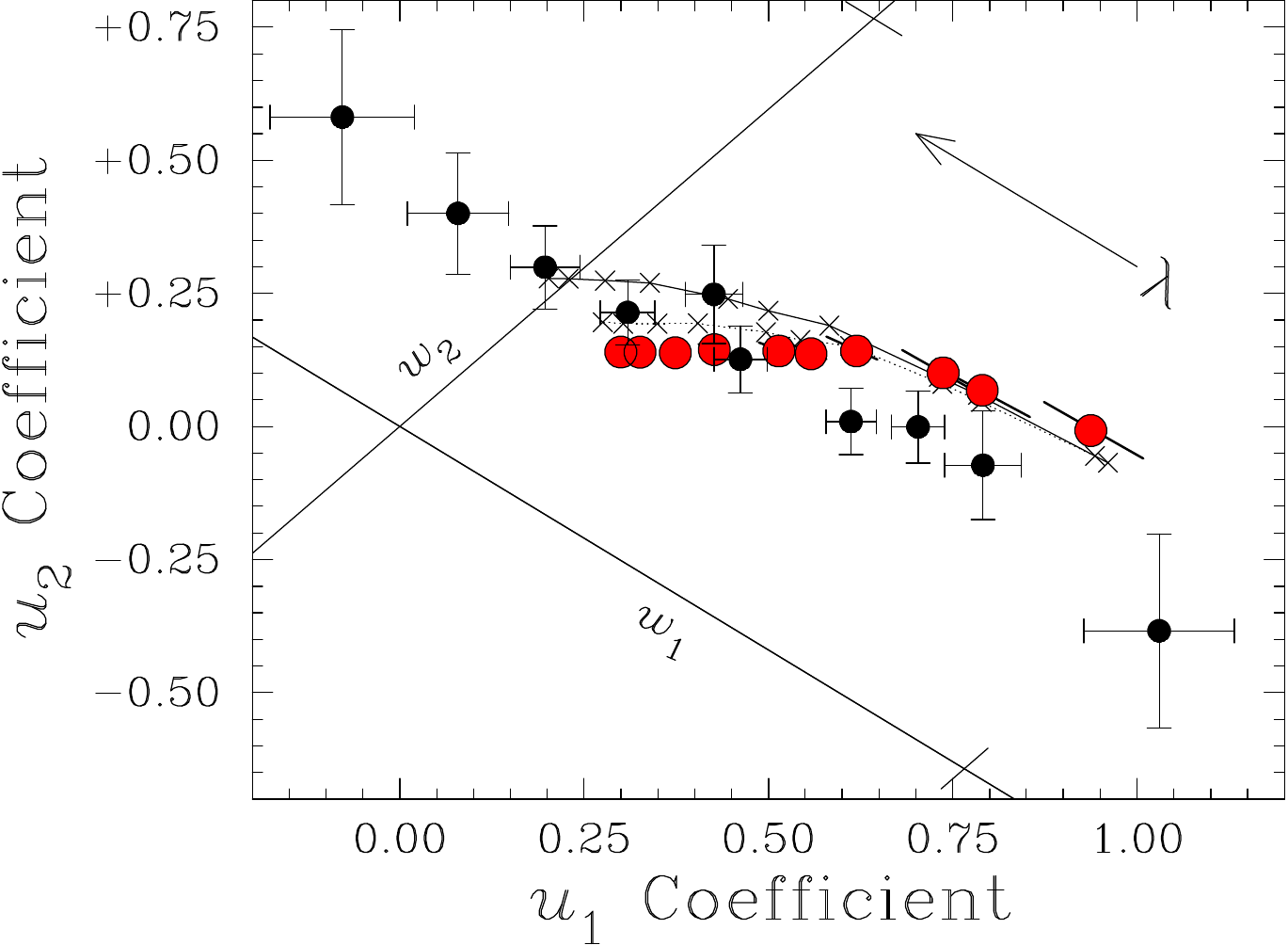}} 
\caption{Comparison of quadratic limb-darkening coefficients for HD~209458.
Black points with error bars, empirical values from
\citet{Southworth08}; red points, SPAM results
(Section~\ref{sec:HDX}).
Continuous and dotted lines
  connect LS2 and FC1 model-atmosphere results, respectively.}
\label{fig:HD2}
\end{figure}

\section{Inferences from exotransit photometry}
\label{sec:EXOX}

The dispersion in coefficient values introduced simply by numerical
techniques poses the question: which procedure is most appropriate for
comparing model-atmosphere results with observational determinations
of limb darkening?  To answer this question it is necessary first to
examine just what it is that is measured from transit observations.

Photo\-metric observations of exoplanetary transits record,
essentially, the variation of $I_\lambda(r)/F_\lambda$ along a chord.
In practice, this variation is parameterized by an analytical
limb-darkening law, whose coefficients are optimized as part of the
global fitting procedure.  This optimization process is quite
different from fitting a limb-darkening law to model-atmosphere
intensities, so it is immediately clear that there cannot be any
simple one-to-one correspondence between photometric and
model-atmopshere LDCs.

Moreover, the extent to which a transit light-curve encodes the global
limb darkening must depend on the impact parameter;\footnote{The
  impact parameter is $p = (a\cos{i})/R_*$ for a circular orbit, where
  $a$ is the orbital semi-major axis and $i$ is the orbital
  inclination} on how faithfully the chosen parametrization of the
limb darkening matches the intensities on those parts of the star that
are occulted; and how well it extrapolates to those parts that are
not.  Given that simple linear and quadratic limb-darkening laws give
only approximate representations of actual intensity distributions, it
might be anticipated that LD coefficients determined from light-curves
must, at some level, suffer systematic biasses depending on impact
parameter, reinforcing the point that these photometric coefficients
must fail to correspond directly to any equivalent, single-valued,
characterization of model-atmosphere results.

To demonstrate and quantify these effects, model transit light-curves
were first generated for a range of stellar temperatures and
passbands.  Limb darkening was represented in the light-curve
calculations by eqtn.~\ref{eq:03}, with coefficients determined by
least squares (i.e., for practical purposes, the model-atmosphere
results were represented almost exactly).  The geometry was set by
choosing a ratio of planetary to stellar radii of 1:10 and a
centres-of-mass separation of 10$R_*$, representative of `hot Jupiter'
systems commonly observed to transit (although the subsequent
limb-darkening results are insensitive to the adopted values), with
the impact parameter varied over the range 0--0.95.

The resulting light-curves were then solved for basic
geometrical parameters, and for linear or quadratic LDCs.  These
calculations were performed using a modified version of {\sc{jktebop}}
\citep{Southworth04a, Southworth04b}, which is itself based on Etzel's
{\sc{ebop}} code \citep{Etzel81, Popper81}.   Proximity effects (tidal
distortion, `reflection', etc.) were throughout assumed to be negligible.

\subsection{Linear law}
\label{sec:llx}

Figure~\ref{fig:uComp} shows a selection of the results, and
demonstrates that, in this parameter space, linear LDCs derived from
photometry systematically increase with increasing impact parameter,
by up to $\sim$0.2 ($\sim$60\%).  This behaviour is straightforward to
understand: in the optical regime $\partial{I}/\partial{\mu}$ decreases
with increasing $\mu$, so that any transit that is not central (i.e.,
inclination $i \ne 90^\circ$) samples a relatively steep part of the
limb-darkening law.  A linear approximation to that law must therefore
yield a linear limb-darkening coefficient that is, in general, larger
than that derived from a central transit.

Figure~\ref{fig:uPlot} illustrates this point by showing
model-atmosphere results, and linear approximations, for a 4~kK model.
The intensity in the \textit{U} band is very nearly a linear function
of $\mu$, so any characterization will yield similar numerical values
for the $u$ coefficient.  This is confirmed both in that least-squares
and flux-conserving approaches yield similar results in this passband,
and in that the $u$ coefficient derived photo\-metrically is insensitive
to impact parameter (upper-left panel of Fig.~\ref{fig:uComp}).  This
contrasts with \textit{H}-band results; the intensity there is a strongly
non-linear function of angle, and any diagnostic that characterizes only small $\mu$
values must yield a larger $u$ coefficient than that characterizing
the entire centre-to-limb variation.  Fig.~\ref{fig:uComp} shows this
to be the case.

Fig.~\ref{fig:uComp} also shows the linear limb-darkening coefficients
obtained by fitting the input intensity distributions directly, using
flux-conserving (FC1) and flux-conserving least-squares (FC2)
techniques, which bracket the range of numerical values derived
directly from the model atmospheres.  In this parameter space, these
coefficients also almost bracket the corresponding photo\-metric LDCs
which suggests a simple, if rough-and-ready, means of comparing
observational and model-atmosphere parametrizations.  Furthermore, if
one had to choose a \textit{single,} linear limb-darkening coefficient to
compare with photo\-metric results, then the FC1 value is probably the
least poor option; for randomly inclined orbits, smaller impact
parameters are more probable than larger ones (and observational
selection effects also favour higher orbital inclinations), and
photo\-metrically determined limb darkening coefficients are generally
closest to the FC1 LDC in this case.\footnote{This conclusion is
  supported by results from many more synthetic light-curves than are
  reported on here.}

\subsection{Quadratic law}
\label{sec:rot1}

The variations in linear limb-darkening coefficient are present
\textit{a fortiori} for the quadratic coefficients, although here the
interpretation is less straightforward because of the well-known strong
correlation between $u_1$ and $u_2$, evident in the
left-hand panels of Fig.~\ref{fig:qPlot} (see also Fig.~2 in \citealt{Southworth08}).
\citet{Pal08} and \citet{Kipping11a}
point out that this correlation is largely removed through a rotation
onto new principal axes,
\begin{align}
\notag
w_1 &= u_1\cos\phi - u_2\sin\phi,\\
w_2 &= u_2\cos\phi + u_1\sin\phi,
\label{eq:w1w2}
\end{align}
with $\phi \simeq 40^\circ$.   Results rotated to these co-ordinates are shown in
the right-hand panel of Fig.~\ref{fig:qPlot}, and confirm that, while
the $w_1$ values continue to show a large variation with impact parameter,
$w_2$ is more nearly constant.

The correspondence between the photo\-metric and
model-atmosphere results is less straightforward than with the linear law; the
model-atmosphere representations of limb darkening show no simple
relationship to the photo\-metrically-determined equivalents
(notwithstanding rough quantitative similarities).   Nevertheless, the
small dispersion found for the $w_2$ coefficient in \textit{both}
observational and model-atmosphere characterizations of limb darkening
indicates that this should be the parameter of choice when making comparisons.

%
%
%
%
%
%

\section{Comparing model-atmosphere and photo\-metric limb-darkening}
\label{sec:LfL}

The foregoing sections emphasize that different characterizations of
model-atmosphere results can give quite different numerical results
(e.g., Fig.~\ref{fig:Rcomp}); and that light-curve analyses, using, of
necessity, approximate limb-darkening `laws', yield LDCs that vary
with transit geometry (e.g., Fig.~\ref{fig:uComp}).  
Furthermore, although a given analytical limb-darkening law is
adopted in photometric studies, the determination of its
coefficients through light-curve modelling is, numerically,
fundamentally distinct from the techniques of fitting
model-atmosphere intensity distributions discussed in
Section~\ref{sec:CLD}; that is, comparing photometric and model-atmosphere
results is, to an extent, like comparing apples and oranges
(but see \citealt{Sandford95, Barone00}).
In order to
examine the relationship between empirical, photo\-metric LDCs and
theoretical model-atmosphere values, it is therefore necessary to
devise a method ensuring a fair comparison.

The most direct way to perform such a `like for like' comparison is
to adapt the methods used in Section~\ref{sec:EXOX}, i.e., to generate
model light-curves for well-studied systems, using as inputs the
empirically determined geometric parameters, coupled to
model-atmosphere intensity distributions (in practice, approximated by
eqtn.~\ref{eq:03}) for the `known' stellar parameters.  This synthetic
photometry can then be solved for the geometric parameters and LDCs,
using the same simplified limb-darkening law adopted in the observational
photo\-metric analysis.  The resulting hybrid
synthetic-photometry/atmosphere-model (SPAM) LDCs
can reasonably be compared directly with empirical values.\footnote{Using a
  simplified description of limb darkening in the fitting step drives the inferred
  geometric parameters away from their input values, but by usually
  unimportant amounts (Appendix~\ref{app:SGP}).}

This approach has been used to investigate two illustrative datasets:
the eight stars with \textit{Kepler} data analysed by
\citet{Kipping11a, Kipping11b}, and the
multiwavelength study of HD~209458 by \citeauthor{Knutson07}
(\citeyear{Knutson07}; see also \citealt{Southworth08}, \citealt{Claret09}).
Synthetic light-curves were generated with 1000 data points through
transit (phases $\pm$0.05). Statistical errors are not quoted on any
results because the analysis is essentially deterministic.

\subsection{\textit{Kepler} targets}

\citet{Kipping11a, Kipping11b} derived quadratic limb-darkening
coefficients for the eight \textit{Kepler} targets they studied.  Their
Monte-Carlo Markov-Chain results are reproduced here in
Fig.~\ref{fig:P6u}.  

Custom model atmospheres were computed for each system as described in
Section~\ref{sec:onf}; adopted stellar parameters are summarized in
Table~\ref{tab:A1} (Appendix~\ref{app:FR}).  This group of stars samples a fairly small range
in atmospheric properties ($\Teff = 5647$:$6297$~K, $\logg = 3.96$:$4.59$,
$\MH = -0.55$:$+0.33$), which is reflected in a rather small range in
model-atmosphere and SPAM LDCs.  It's therefore somewhat surprising
that agreement between empirical LDCs and those from the SPAM approach
(or from direct fitting to model atmospheres\footnote{In general, the
  SPAM results are closest to the FC1 direct characterization of
  intensities.})  varies from excellent (Kepler-6) to statistically
unacceptable (e.g., Kepler-5).

There is a suggestion that the extent of agreement correlates with
temperature; the SPAM LDCs for the three coolest stars fall within the
cloud of the best-fitting 90\%\ of solutions, while those for the
three hottest lie (just) outside.\footnote{The cooler stars in this
  sample are also those with higher gravities and metallicities, so
  temperature is not necessarily the key parameter.}  The trend is for
the cloud of empirical values to move towards smaller ($u_1,u_2$)
values with increasing temperature, compared to the model-atmosphere
results.  It's unclear why the empirical results should show so much
greater variation than the models, suggesting that this apparent trend
may simply be an artefact of the small sample, or that some additional
factor plays an unexpectedly important role.

\subsection{HD 209458}
\label{sec:HDX}

Baseline parameters of $\Teff = 6113$~K \citep{Casagrande10}, $\logg =
4.50$, $\MH = +0.03$ \citep{Sousa08}, $\vt = 2$~\kms, $\lH = 1.25$
were adopted to construct the reference model atmosphere and
intensities for HD~209458.  
Broad-band limb-darkening was
calculated by assuming `top hat' response functions for the
photo\-metric passbands of the \citeauthor{Knutson07} HST observations.
The principal results are summarized in Table~\ref{tab:A2}

Additional models were run for $\Teff = 5913, 6313$; $\logg = 4.2,
4.8$; $\lH = 0.5$; $\vt = 0, 4~\kms$; and $\MH = -0.4, +0.4$.  These
ranges allow for quite generous uncertainties in parameters for this
well-studied system. The extremes in linear LDCs from the models are
for the low-\Teff\ and high-gravity models (numerically largest and
smallest coefficients, respectively), and these models are used to
illustrate plausible `error bars' on the SPAM coefficients
in Figs.~\ref{fig:HD1} and~\ref{fig:HD2}.

Fig.~\ref{fig:HD1} shows results for linear coefficients.  The
discrepancies between model-atmosphere and photo\-metric results already
noted
by \citeauthor{Claret09} (\citeyear{Claret09}; see also
\citealt{Southworth08}), on the basis of older models,\footnote{The
  principal cause of the minor quantitative differences between the
  results shown in Fig.~\ref{fig:HD1} appears to be the different
  treatments of convection adopted here and by \citeauthor{Kurucz93} 
(\citeyear{Kurucz93};   the
  source of Claret's models).}  persist in the new analysis.

The comparison for quadratic coefficients is shown in
Fig.~\ref{fig:HD2}.
The
  variation with wavelength of both $u_1$ and $u_2$ coefficients is
  much less for the SPAM coefficients than is found
  empirically. However, both sequences run almost parallel to the
  rotated $w_1$ axis, and agreement in the better-determined $w_2$
  parameter is tolerable at all wavelengths.
\footnote{The observational uncertainties can't be straightforwardly
  propagated from $\sigma(u_{1,2})$ to $\sigma(w_{1,2})$ because of
  the strong correlation between $u_1$ and $u_2$.  It should also be
  noted that the error bars in
  Fig.~\ref{fig:HD2} represent the 68\%\ dispersion in parameter
  values obtained from MC replications (Southworth, personal
  communication); they therefore can't be compared directly to the
  dispersion in results shown in Fig.~\ref{fig:P6u}, which 
show parameters from the 90/95\%\ of solutions with the smallest
overall $\chi^2$ values.}
In particular, for the $\sim$678~nm passband, which is close to the effective
wavelength of the \textit{Kepler} results, the agreement is reasonably good,
[$(w_1,w_2) = (0.234,0.385), (0.099,0.363)$ for SPAM and light-curve
coefficients, respectively].
This is in contrast to the \textit{Kepler} results for stars at similar
effective temperatures (but is consistent with the result that it
is the higher-gravity stars that show the best agreement between
models and observations).





\section{Summary and conclusions}

Different methods of fitting a given limb-darkening law to a given
model-atmosphere intensity distribution lead to quite different
numerical coefficients.  Furthermore, the limb-darkening coefficients
determined from photometry of exoplanetary transits are functions of
impact parameter, and can't reliably be compared directly to
any of the standard model-atmosphere characterizations.

A more direct comparison can be made if the model intensities are
translated into observer space, through the medium of synthetic
light-curves.  The resulting synthetic-photometry/atmosphere-model
(SPAM) limb-darkening coefficients are not single-valued, but
\textit{can} be compared directly with empirical results.  

[If one had to choose a traditional single-valued representation of
model-atmosphere results, then at optical wavelengths closest
agreement with the SPAM results is generally obtained with the
standard (FC1) flux-conserving method, which also yields the smallest
value of the linear limb-darkening coefficient.]

For the commonly used quadratic limb-darkening law, most of the
variation in different fits to model-atmosphere intensities is in the $w_1$
parameter, with much smaller dispersions in the $w_2$ coefficient.
Since the $w_1$ axis is also defined as that which maximizes
dispersion in observational (Monte-Carlo) results, the most sensitive
comparison between models and observations is in $w_2$.

New model-atmosphere calculations, analysed with the SPAM approach,
show mixed results.  Agreement with empirical \textit{Kepler}
LDCs is good in some cases (differences in $w_2$ 
less than 0.06 in four out of six systems), but not in others.  There is a hint of a
possible temperature dependence in the extent of disagreement for
these targets, with cooler stars showing better agreement.
However, at similar effective wavelengths
HST results for HD~209458 (which is at the hotter end of the range of
\textit{Kepler} targets) agree well with models;  there are
discrepancies at longer and short wavelengths, though again
with fair agreement in $w_2$.  Since gravity (and metallicity)
correlate with temperature for the \textit{Kepler} sample, and since HD~209458 is
both high-temperature and high-gravity in the context of that sample,
this might be taken as an indication that agreement is better at
higher gravities (with temperature as a secondary factor).  However, the
ranges in all quantitities characterizing the atmospheres are so small
as to render such conclusions speculative at this stage.
Forthcoming results from \textit{Kepler}, and other missions, should enlarge the
parameter space, and permit better discrimination of where models and
observations do and do not agree.

%

\section*{Acknowledgments}

I'm grateful to John Southworth for making {\sc{jktebop}} available, and for
perceptive comments on the manuscript; and to David Kipping, for
instructive conversations, and for providing the MCMC results used in
Figure~\ref{fig:P6u}.  I also thank Luca Casagrande for a
helpful and constructive referee's report.

\appendix
\section{Fit results}
\label{app:FR}
\begin{table*}
  \caption{Limb-darkening results:  \textit{Kepler} photometry.  For each star,
    the physical parameters used in the model-atmosphere calculations
    are first listed,
    followed by the resulting 4-parameter \textit{Kepler}-band limb-darkening coefficients (eqtn.~\ref{eq:03}).
    Subsequent columns list the SPAM coefficients, determined by fitting
    synthetic light-curves generated from the `known' system parameters;  
    the empirical photo\-metric
    coefficients;  and, for reference, coefficients determined directly
    from the model-atmosphere intensity distributions.   All necessary
    stellar
    \& system parameters are adopted from \citet{Kipping11a, Kipping11b}.
  }
\begin{tabular}{lccccccc}
\hline
      & Photometry    & Photometry   &
      \multicolumn{4}{c}{\hrulefill{ Model-atmosphere fits }\hrulefill} \\
      & (Synthetic)   & (Observed) & LS1 & LS2 & FC1 & FC2 \\
\hline
\hline
\multicolumn{1}{l}{Kepler-4}&
\multicolumn{6}{l}{$\Teff = 5857$~K, $\logg = 4.25$, $\MH = +0.17$, $\vt = 2$~\kms, $\lH = 1.25$.}\\
\multicolumn{1}{l}{$a_n, n=1,4$}&
+7.63788E-01 & $-$7.97285E-01 & +1.40090E+00 & $-$5.86378E-01 \\
\hline
linear, $u$  & +0.6080       &            & +0.6252   & +0.6491   & +0.5828   & +0.6491   \\
quad, $u_1$  & +0.5201       &$+0.61^{+0.59}_{-0.39}$
                                          & +0.4805   & +0.4480   & +0.5035   & +0.4481   \\
quad, $u_2$  & +0.1230       &$-0.21^{+0.52}_{-0.68}$
                                          & +0.1931   & +0.2256   & +0.1586   & +0.2256   \\
\hline
\hline

\multicolumn{1}{l}{Kepler-5}&
\multicolumn{6}{l}{$\Teff = 6297$~K, $\logg = 3.96$, $\MH = +0.03$, $\vt = 2$~\kms, $\lH = 1.25$.}\\
\multicolumn{1}{l}{$a_n, n=1,4$}&
+7.34699E-01 & $-$7.94487E-01 & +1.43212E+00 & $-$6.20527E-01 \\
\hline
linear, $u$  & +0.5564       &            & +0.6011   & +0.6265   & +0.5566   & +0.6267   \\
quad, $u_1$  & +0.4877       &$+0.25^{+0.13}_{-0.12}$
                                          & +0.4537   & +0.4248   & +0.4739   & +0.4250   \\
quad, $u_2$  & +0.1377       &$+0.37^{+0.25}_{-0.27}$
                                          & +0.1966   & +0.2254   & +0.1654   & +0.2253   \\
\hline
\hline

\multicolumn{1}{l}{Kepler-6}&
\multicolumn{6}{l}{$\Teff = 5647$~K, $\logg = 4.59$, $\MH = +0.33$, $\vt = 2$~\kms, $\lH = 1.25$.}\\
\multicolumn{1}{l}{$a_n, n=1,4$}&
+8.20192E-01 & $-$9.18046E-01 & +1.53037E+00 & $-$6.26480E-01 \\
\hline
linear, $u$  & +0.5967       &            & +0.6436   & +0.6664   & +0.6025   & +0.6665   \\
quad, $u_1$  & +0.5415       &$+0.55^{+0.13}_{-0.11}$
                                          & +0.5011   & +0.4675   & +0.5262   & +0.4675   \\
quad, $u_2$  & +0.1189       &$+0.01^{+0.26}_{-0.27}$
                                          & +0.1901   & +0.2240   & +0.1527   & +0.2240   \\
\hline
\hline

\multicolumn{1}{l}{Kepler-7}&
\multicolumn{6}{l}{$\Teff = 5933$~K, $\logg = 3.98$, $\MH = +0.11$, $\vt = 2$~\kms, $\lH = 1.25$.}\\
\multicolumn{1}{l}{$a_n, n=1,4$}&
+7.48602E-01 & $-$7.78844E-01 & +1.38072E+00 & $-$5.77728E-01 \\
\hline
linear, $u$  & +0.5850       &            & +0.6205   & +0.6439   & +0.5790   & +0.6440   \\
quad, $u_1$  & +0.5191       &$+0.34^{+0.16}_{-0.13}$
                                          & +0.4795   & +0.4480   & +0.5017   & +0.4481   \\
quad, $u_2$  & +0.1183       &$+0.33^{+0.26}_{-0.34}$
                                          & +0.1881   & +0.2196   & +0.1546   & +0.2195   \\
\hline
\hline

\multicolumn{1}{l}{Kepler-8}&
\multicolumn{6}{l}{$\Teff = 6213$~K, $\logg = 4.28$, $\MH = -0.55$, $\vt = 2$~\kms, $\lH = 1.25$.}\\
\multicolumn{1}{l}{$a_n, n=1,4$}&
+7.13647E-01 & $-$7.27611E-01 & +1.35455E+00 & $-$5.92941E-01 \\
\hline
linear, $u$  & +0.5817       &            & +0.5980   & +0.6240   & +0.5525   & +0.6241   \\
quad, $u_1$  & +0.4864       &$+0.41^{+0.55}_{-0.25}$
                                          & +0.4474   & +0.4176   & +0.4675   & +0.4179   \\
quad, $u_2$  & +0.1337       &$+0.11^{+0.44}_{-0.83}$
                                          & +0.2008   & +0.2305   & +0.1701   & +0.2304   \\
\hline
\hline

\multicolumn{1}{l}{TrES-2}&
\multicolumn{6}{l}{$\Teff = 5850$~K, $\logg = 4.40$, $\MH = +0.15$, $\vt = 2$~\kms, $\lH = 1.25$.}\\
\multicolumn{1}{l}{$a_n, n=1,4$}&
+6.97192E-01 & $-$6.67832E-01 & +1.29178E+00 & $-$5.62467E-01 \\
\hline
linear, $u$  & +0.6238       &            & +0.6117   & +0.6366   & +0.5676   & +0.6368   \\
quad, $u_1$  & +0.4754       &$+0.52^{+0.44}_{-0.34}$
                                          & +0.4644   & +0.4344   & +0.4846   & +0.4346   \\
quad, $u_2$  & +0.1608       &$+0.06^{+0.37}_{-0.48}$
                                          & +0.1964   & +0.2264   & +0.1659   & +0.2264   \\
\hline
\hline
\end{tabular}
\label{tab:A1}
\end{table*}

\begin{table*}
\caption{Limb-darkening results:  HD 209458.   System  parameters
and observed LDCs are from \citeauthor{Southworth08} (\citeyear{Southworth08}).
Stellar parameters
$\Teff = 6113$~K \citep{Casagrande10}, 
$\logg = 4.50$, $\MH = +0.03$ \citep{Sousa08}, $\vt = 2$~\kms, $\lH =
1.25$.   Columns follow the model of Table~\ref{tab:A1}, with rows
grouped by HST wavelength in nm.}
\begin{tabular}{lccccccc}
\hline
      & Photometry    & Photometry   &
      \multicolumn{4}{c}{\hrulefill{ Model-atmosphere fits }\hrulefill} \\
      & (Synthetic)   & (Observed) & LS1 & LS2 & FC1 & FC2 \\
\hline
\hline
HST-320;&
$a_n, n=1,4$ &  +4.58205E-01 & -7.02251E-01 & +1.97519E+00 & -7.98143E-01\\
\hline
linear, $u$ & $+0.9346$ & $+0.828{\pm}0.023$ & $+0.9064$ & $+0.9029$ & $+0.9151$ & $+0.9029$ \\
quad, $u_1$ & $+0.9373$ & $+1.030{\pm}0.102$ & $+0.9438$ & $+0.9607$ & $+0.9428$ & $+0.9607$ \\
quad, $u_2$ & $-0.0076$ & $-0.384{\pm}0.182$ & $-0.0500$ & $-0.0681$ & $-0.0553$ & $-0.0681$ \\
\hline
\hline
HST-375;&
$a_n, n=1,4$ &  +6.43615E-01 & -9.36895E-01 & +2.09622E+00 & -8.88657E-01\\
\hline
linear, $u$ & $+0.8204$ & $+0.754{\pm}0.013$ & $+0.8283$ & $+0.8356$ & $+0.8118$ & $+0.8356$ \\
quad, $u_1$ & $+0.7901$ & $+0.791{\pm}0.052$ & $+0.7809$ & $+0.7844$ & $+0.7888$ & $+0.7844$ \\
quad, $u_2$ & $+0.0680$ & $-0.073{\pm}0.012$ & $+0.0632$ & $+0.0596$ & $+0.0460$ & $+0.0596$ \\
\hline
\hline
HST-430;&
$a_n, n=1,4$ &  +6.15746E-01 & -8.44919E-01 & +2.00870E+00 & -8.87838E-01\\
\hline
linear, $u$ & $+0.7839$ & $+0.703{\pm}0.007$ & $+0.8005$ & $+0.8118$ & $+0.7758$ & $+0.8118$ \\
quad, $u_1$ & $+0.7370$ & $+0.703{\pm}0.036$ & $+0.7283$ & $+0.7312$ & $+0.7360$ & $+0.7312$ \\
quad, $u_2$ & $+0.1006$ & $-0.001{\pm}0.068$ & $+0.0964$ & $+0.0933$ & $+0.0797$ & $+0.0933$ \\
\hline
\hline
HST-485;&
$a_n, n=1,4$ &  +6.33987E-01 & -6.21291E-01 & +1.55033E+00 & -7.10907E-01\\
\hline
linear, $u$ & $+0.6919$ & $+0.618{\pm}0.006$ & $+0.7276$ & $+0.7482$ & $+0.6865$ & $+0.7483$ \\
quad, $u_1$ & $+0.6197$ & $+0.612{\pm}0.034$ & $+0.5962$ & $+0.5826$ & $+0.6109$ & $+0.5826$ \\
quad, $u_2$ & $+0.1418$ & $+0.009{\pm}0.062$ & $+0.1754$ & $+0.1894$ & $+0.1511$ & $+0.1894$ \\
\hline
\hline
HST-540;&
$a_n, n=1,4$ &  +7.10040E-01 & -7.19161E-01 & +1.47157E+00 & -6.49635E-01\\
\hline
linear, $u$ & $+0.6307$ & $+0.561{\pm}0.007$ & $+0.6684$ & $+0.6919$ & $+0.6248$ & $+0.6919$ \\
quad, $u_1$ & $+0.5578$ & $+0.426{\pm}0.039$ & $+0.5240$ & $+0.4999$ & $+0.5437$ & $+0.5000$ \\
quad, $u_2$ & $+0.1364$ & $+0.248{\pm}0.092$ & $+0.1927$ & $+0.2171$ & $+0.1622$ & $+0.2171$ \\
\hline
\hline
HST-580;&
$a_n, n=1,4$ &  +7.20922E-01 & -6.88211E-01 & +1.34528E+00 & -5.92544E-01\\
\hline
linear, $u$ & $+0.5917$ & $+0.534{\pm}0.006$ & $+0.6322$ & $+0.6583$ & $+0.5852$ & $+0.6584$ \\
quad, $u_1$ & $+0.5141$ & $+0.462{\pm}0.036$ & $+0.4756$ & $+0.4453$ & $+0.4967$ & $+0.4455$ \\
quad, $u_2$ & $+0.1417$ & $+0.126{\pm}0.063$ & $+0.2089$ & $+0.2394$ & $+0.1770$ & $+0.2393$ \\
\hline
\hline
HST-678;&
$a_n, n=1,4$ &  +7.67414E-01 & -7.43741E-01 & +1.22290E+00 & -5.25063E-01\\
\hline
linear, $u$ & $+0.5090$ & $+0.437{\pm}0.006$ & $+0.5520$ & $+0.5824$ & $+0.5011$ & $+0.5826$ \\
quad, $u_1$ & $+0.4267$ & $+0.309{\pm}0.037$ & $+0.3799$ & $+0.3392$ & $+0.4044$ & $+0.3395$ \\
quad, $u_2$ & $+0.1442$ & $+0.214{\pm}0.061$ & $+0.2295$ & $+0.2696$ & $+0.1935$ & $+0.2695$ \\
\hline
\hline
HST-775;&
$a_n, n=1,4$ &  +7.82449E-01 & -7.91226E-01 & +1.16704E+00 & -4.88230E-01\\
\hline
linear, $u$ & $+0.4545$ & $+0.377{\pm}0.008$ & $+0.4964$ & $+0.5282$ & $+0.4458$ & $+0.5286$ \\
quad, $u_1$ & $+0.3737$ & $+0.197{\pm}0.047$ & $+0.3240$ & $+0.2785$ & $+0.3494$ & $+0.2790$ \\
quad, $u_2$ & $+0.1384$ & $+0.299{\pm}0.078$ & $+0.2299$ & $+0.2741$ & $+0.1929$ & $+0.2739$ \\
\hline
\hline
HST-873;&
$a_n, n=1,4$ &  +7.95398E-01 & -8.55463E-01 & +1.17721E+00 & -4.90744E-01\\
\hline
linear, $u$ & $+0.4088$ & $+0.324{\pm}0.011$ & $+0.4498$ & $+0.4837$ & $+0.3996$ & $+0.4838$ \\
quad, $u_1$ & $+0.3259$ & $+0.079{\pm}0.069$ & $+0.2751$ & $+0.2289$ & $+0.3030$ & $+0.2290$ \\
quad, $u_2$ & $+0.1396$ & $+0.400{\pm}0.114$ & $+0.2331$ & $+0.2774$ & $+0.1930$ & $+0.2773$ \\
\hline
\hline
HST-971;&
$a_n, n=1,4$ &  +7.63034E-01 & -7.89085E-01 & +1.07318E+00 & -4.49903E-01\\
\hline
linear, $u$ & $+0.3835$ & $+0.275{\pm}0.016$ & $+0.4242$ & $+0.4588$ & $+0.3732$ & $+0.4591$ \\
quad, $u_1$ & $+0.2997$ & $-0.078{\pm}0.098$ & $+0.2496$ & $+0.2021$ & $+0.2753$ & $+0.2025$ \\
quad, $u_2$ & $+0.1398$ & $+0.581{\pm}0.164$ & $+0.2330$ & $+0.2782$ & $+0.1958$ & $+0.2780$ \\
\hline
\hline
\end{tabular}
\label{tab:A2}
\end{table*}

\section{Systematics of geometric parameters}
\label{app:SGP}

\begin{figure*}
\center{\includegraphics[scale=0.9,angle=0]{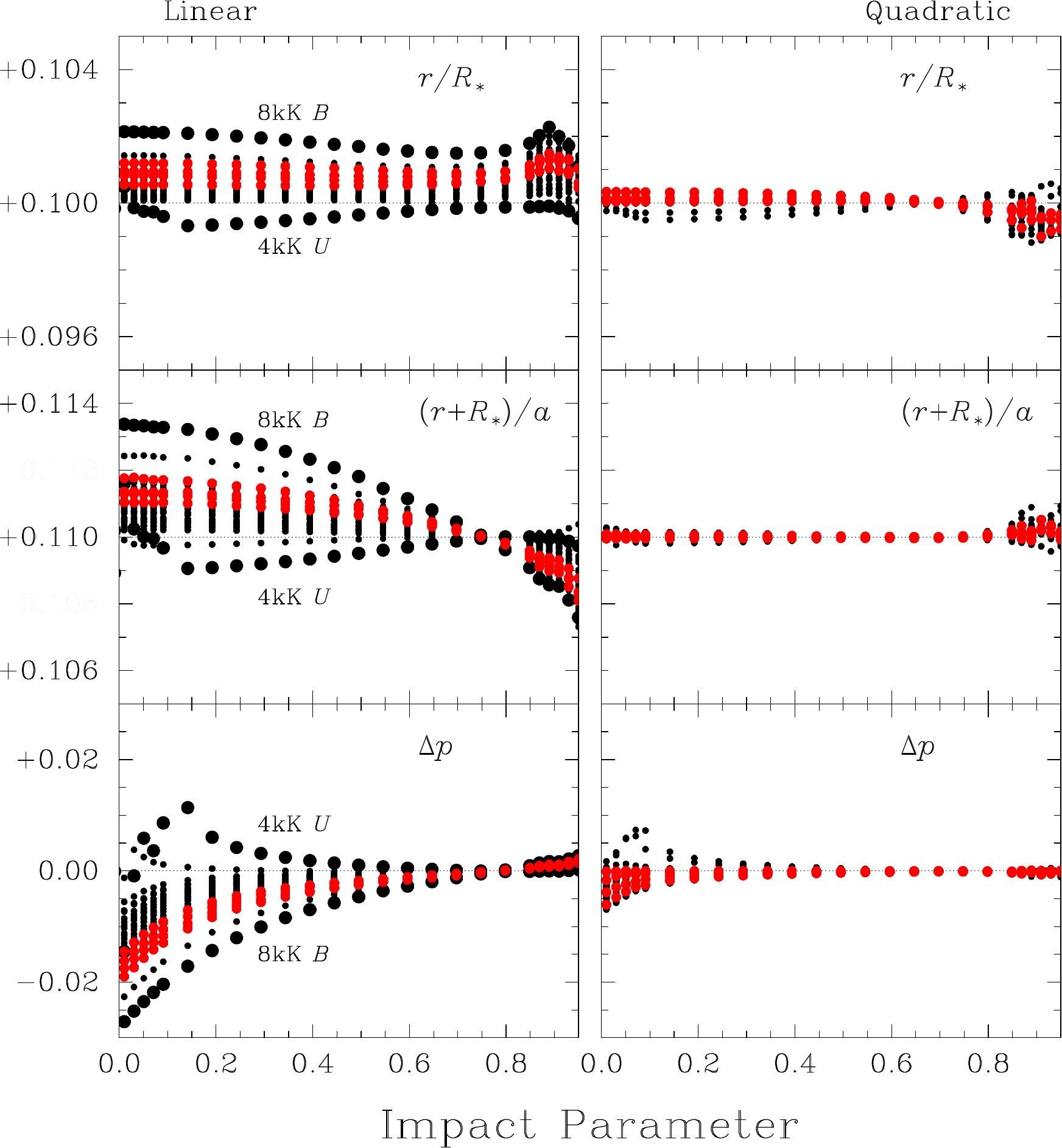}} 
\caption{Geometric parameters determined from modelling synthetic
  light-curves, as functions of input impact parameter.  The
  light-curves were generated using essentially exact representations
  of limb darkening; ratio of planetary to stellar radii $r/R_* =
  0.1$; the sum of the radii, in units of the semi-major axis
  $r+R_*)/a = 0.11$.  Light-curve solutions assume linear (left-hand
  panels) or quadratic (right-hand panels) limb darkening, with
  coefficients optimised as part of the fitting process.  The
  (generally small) departures from input parameters are solely a
  result of using these approximate representations of limb darkening.
  Results are shown for stellar models at 4, 6, 8, and 12~kK, and
\textit{UBVRIJHKL} passbands;  4-kK \textit{U}-band and 8-kK
\textit{B}-band results  bracket most of the data when  a linear
law is adopted, and are highlighted with larger symbols.  Results for the
\textit{R}~band,
which are representative of typical unfiltered CCD systems, are shown
in red.}
\label{fig:app1}
\end{figure*}

Because light-curves have only a rather weak dependence on limb
darkening, we might expect that the use of simplified limb-darkening
laws, or even moderately inaccurate LDCs, should have only a very modest
effect on the determination of basic geometric parameters when
modelling photometry.  To demonstrate this (at the referee's
suggestion), additional results from the model grids described in
Section~\ref{sec:EXOX} are presented in Fig~\ref{fig:app1}.

[To remind the reader, model light-curves were generated using
a ratio of planetary to stellar radii of 1:10 and a centres-of-mass
separation of 10$R_*$, over a range of impact parameters, and an
effectively exact representation of limb darkening.   These
light-curves were then solved, adopting linear or quadratic limb
darkening (with the LDCs as free parameters).   It is the results of
these light-curve solutions that are summarized in
Fig~\ref{fig:app1}.]

The systematic errors in fitted geometric parameters (up to $\sim$4\%\
in $r/R_*$ and $(r+R_*)/a$ for the models presented here) might be
significant for the best-quality photometry if a linear limb-darkening
law is assumed, but are negligible if a quadratic law is used.

\bsp

\label{lastpage}

\end{document}